\documentclass[aip,pop]{revtex4-1}

\usepackage{color}
\usepackage{graphicx}
\usepackage{subfigure}
\usepackage{epstopdf}
\usepackage{amsmath}
\usepackage{dcolumn}
\usepackage{bm}
\usepackage[colorlinks=true, linkcolor=blue]{hyperref}

\begin{document}

\title{Efficiency of Wave-Driven Rigid Body Rotation Toroidal Confinement}

\author{J.~M.~Rax}
\affiliation{Universit\'{e} de Paris XI - Ecole Polytechnique, LOA-ENSTA-CNRS, 91128 Palaiseau, France}
\author{R.~Gueroult}
\affiliation{LAPLACE, Universit\'{e} de Toulouse, CNRS, INPT, UPS, 31062 Toulouse, France}
\author{N.~J.~Fisch}
\affiliation{Department of Astrophysical Sciences, Princeton University, Princeton, New Jersey 08544, USA}

\begin{abstract}
The compensation of vertical drifts in toroidal magnetic fields through a wave-driven poloidal rotation is compared to compensation through the wave driven toroidal current generation to support the classical magnetic rotational transform. 
The advantages and drawbacks associated with the sustainment of a radial electric field are compared with those associated with the sustainment of a poloidal magnetic field both in terms of energy content and power dissipation. 
The energy content of a radial electric field is found to be smaller than the energy content of a poloidal magnetic field for a similar set of orbits. 
The wave driven radial electric field generation efficiency is similarly shown, at least in the limit of large aspect ratio, to be larger than the efficiency of wave-driven toroidal current generation.
\end{abstract}
\date{\today}
\maketitle
\endpage{}

\section{Introduction}

In a static homogeneous magnetic field, the orbit of a charged particle is a combination of  rotation around the field lines and a translation along the field lines.
With an inhomogeneous static magnetic field, two general magnetic trap configurations  can be considered: 
The first is the open field line configuration, where the magnetic field lines are closed outside the plasma. 
This configuration must display a minimum of a trapping potential along the open field lines to restrict the parallel motion and to achieve confinement. 
For non-neutral plasma in Penning trap this potential is electrostatic.
For thermonuclear quasi-neutral plasma,  this potential is associated with the diamagnetic force leading to magnetic mirroring.

The second magnetic confinement trap, and the topic of interest here,  features closed field lines, that is to say, a toroidal topology.
We will call $R_{0}$ the major radius of the plasma torus, $a$ the minor radius of
this torus and $B_{0}$ the magnetic field on the magnetic axis.
There is no need to create a minimum of a potential along the field lines,  as particles explore the full length of the line. 
But we have to compensate the magnetic toroidal vertical drift
velocity $v_{D}$ across the field lines, 
\begin{equation}
v_{D}=\frac{{v_{\Vert}}^{2}+{v_{c}}^{2}/2}{R_{0}\omega _{c_0}}\approx \frac{%
2k_{B}T}{eR_{0}B_{0}},  \label{vdrift}
\end{equation}
where $v_{\Vert }$ the velocity along the field lines, $v_{c}$ the cyclotron
velocity around the field lines and $\omega _{c_{0}}=eB_{0}/m$ is the
cyclotron frequency of a particle with charge $e$ and mass $m$, $T$ is the
temperature of the associated population.

There are two ways to compensate this vertical drift:

({\it i}) This compensation can be achieved with the magnetic rotational transform~\cite{Spitzer1958} which  short circuits the vertical drift current associated with Eq.~(\ref{vdrift}),
thus providing steady state confinement. Stellarators and tokamak are the two main configurations designed according to this principle~\cite{Spitzer1958,Budker1961}. We note in passing that, in particular tokamak equilibria referred to as \emph{current hole} configurations, the magnetic rotational transform can be achieved with zero toroidal current and poloidal magnetic field on axis~\cite{Cowley1991,Gourdain2009,Fujita2010}.

({\it ii}) Instead of using a poloidal magnetic field, the short circuiting of the vertical drift current can be achieved through a radial electric field ${\bf E}$, resulting in an $E/B_{0}$ poloidal rotation.  
This configuration also avoids the vertical magnetic escape described by Eq.~(\ref{vdrift}).
%$\left( \text{%
%\ref{vdrift}}\right) $. 
This is the principle of toroidal {\it %
magnetoelectric} confinement~\cite{Budker1961,Stix1970}.
However, because of the difficulty to generate and control a radial electric field inside a hot plasma, electric rotation schemes have been far less explored than magnetic rotational transforms which underlie  the basic field configurations of tokamak and stellarators.

%These three magnetic confinement principles: ({\it i}) open field lines and
%the necessity of a potential along the field lines to avoid the parallel
%escape, and closed field lines and the necessity of a slow ({\it ii})
%magnetic or ({\it iii}) electric rotation across the field lines to
%compensate the vertical escape, have been considered since the early times
%of thermonuclear plasma physics. Because of the difficulty to generate and
%control a radial electric field inside a hot plasma, electric rotation
%schemes have been far less explored than magnetic rotational transforms such
%as tokamak and stellarators.

\begin{figure}
\begin{center}
\includegraphics[width = 8cm]{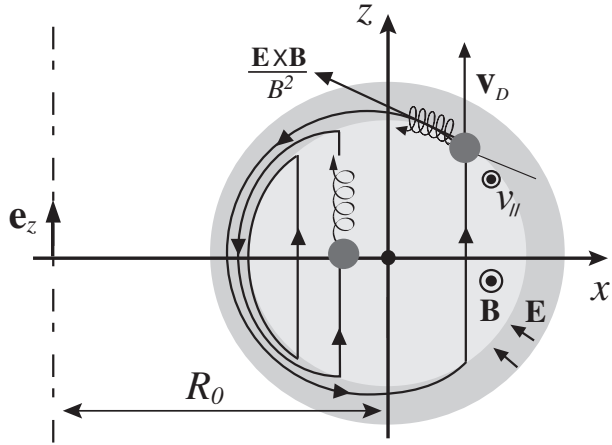}
\caption{Magneto-electric Stix configuration with an edge rotational transform leading to D shaped orbits. }
\label{Fig:Figure1}
\end{center}
\end{figure}

The electric rotation schemes previously considered envisioned rotational transform only near the plasma periphery.
 Four decades ago, T.~H.~Stix identified~\cite{Stix1970}, described and analyzed~\cite{Stix1971,Stix1971a} a toroidal trap based on a purely toroidal magnetic field supplemented by a radial electric field localized near the edge of the plasma and sustained by a preferential loss of fast ions. 
The orbit in such a trap comprises a vertical drift $v_{D}$ near the center of the discharge, closed by an electric $E/B_{0}$ rotation at the edge. 
These {\sf D} shaped orbits, depicted in Fig.~\ref{Fig:Figure1}, are closed by the $E/B_{0}$ rotation on the low field side or on the high field side depending on the sign of the charge. 
Stix studied the self-consistent equilibrium and the main instabilities associated with this original trap~\cite{Stix1971,Stix1971a}. 
The impact of edge biasing was studied experimentally, with edge electrodes, in what was called an electric tokamak program~\cite{Taylor2002,Reece1997,Taylor2005}.
Edge electrodes produced observable effects on plasma confinement in tokamaks \cite{Nascimento05}.
The impact of electric rotation in various toroidal configurations was also  analyzed~\cite{Tsypin1998}. 
Electric rotation to overcome drifts was likewise analyzed, both theoretically and experimentally, within the context of particle
accelerators~\cite{Janes1965,Janes1965a,Janes1966,Daugherty1967} and non-neutral plasma confinement~\cite{Avinash1991,Stoneking2004,Hurricane1998}. 

To produce single-particle confinement, here we explore the possibility to replace essentially completely the poloidal magnetic field (and with it the toroidal current) by a radial electric field (radial polarization) extending from the magnetic axis toward the edge. 
The scheme proposed here is different from the Stix proposal and related configurations in two important respects:
%  ~\cite{Stix1970,Stix1971,Stix1971a} and from electric tokamak configuration~\cite{Taylor2002,Reece1997}: 
%
({\it i}) here we  consider a radial electric field extending  from the center toward the edge rather than just near the edge;
and, relatedly, ({\it ii}),  here we consider a volumetric wave-driven electric field, through angular momentum absorption, rather than relying on preferential ions losses at the edge or edge biasing. We demonstrate that, for the purpose of single particle confinement, steady-state, wave-driven, electric field generation is more efficient than the classical wave-driven toroidal current schemes~\cite{Fisch1978,Fisch1987}, both in terms of energy storage and, at least in the limit of large aspect ratio, power dissipation. We call this new confinement scheme the {\it wave-driven rotating torus}, or, what we refer to more compactly as, the {\it WDRT}.

To estimate the potential of WDRT toroidal confinement, where there is no poloidal magnetic field, we  consider the classical tokamak with a  poloidal magnetic field as a benchmark case. For the same toroidal magnetic field $B_{0}$ and major/minor radius, $R_{0}/a$, we set up a comparative analysis of the confined orbits when the vertical drift is compensated: 
The compensation occurs either ({\it i}) with a poloidal magnetic field (tokamak), or ({\it ii}) with a radial  electric field (WDRT). 
For definiteness, we consider a radial electric field or a poloidal magnetic field increasing linearly with minor radius.   
In either case, such a field produces solid body rotation. By matching a single parameter, we can show that the orbits display the same confinement geometry in both the electric and magnetic cases. This provides a robust comparative criteria between WDRT and tokamak schemes. This criteria will then be used to compare the energy content and the power requirement of both wave-driven steady-state confinement schemes.

The study here is limited to comparing reactive energy storage and active power consumption associated with wave-driven poloidal magnetic field generation versus wave-driven radial electric field generation for orbit confinement. Pressure equilibrium is not addressed here, nor are fluid and kinetic instabilities. These are all critical issues to address. However, it is our hope that the unusual and attractive properties of the WDRT identified here, both in terms of reactive energy storage and power consumption,  will motivate exploration of the important issues not covered here.

The paper is organized as follows:  
In Sec.~\ref{Sec:II}, we identify the single parameter that can be used to compare the energy content and power dissipation for both confinement schemes. 
% We  show that WDRT orbits can be compared to tokamak orbits on the basis of this single parameter.
%
In Sec.~\ref{Sec:III}, we show that closed orbits are possible with small displacement  in the WDRT even for alpha particles.
In Sec.~\ref{Sec:IV}, we compare the energy content associated with producing a magnetic rotational transform to that associated with producing an electric rotational transform. 
In Sec.~\ref{Sec:V}, we review wave-particle dynamics.
In Sec.~\ref{Sec:VI}, for the very same orbital confinement properties, we calculate the power consumption of both the WDRT scheme and the traditional non-inductively driven, steady-state tokamak. 
%We describe and analyze wave-driven radial electric field generation and review wave-driven toroidal current generation~\cite{Fisch1978,Fisch1987}. 
 %
On the basis of these results, in Sec.~\ref{Sec:VII}, we set up a comparative analysis of wave driven poloidal magnetic field generation versus wave driven radial electric field generation, and show
that, at least in the limit of large aspect ratio, the electric case power requirement is smaller than the magnetic case one.
We summarize our findings and conclusions in the last section. 
Appendix~\ref{Sec:appA} provides a more detailed orbit analysis than that given in Sec.~\ref{Sec:II}. 
The geometry of the slowing down orbits of thermonuclear alphas is presented in Appendix~\ref{Sec:appB}.
Appendix~\ref{Sec:appC} reviews the classical wave-particle energy-momentum transfer relations used in Sec.~\ref{Sec:V} to evaluate the efficiency of steady-state angular momentum sustainment with waves.
In Appendix~\ref{Sec:appD}, we address horizontal polarization in WDRT traps.

%\textcolor{red}{  This is how you color it ...
%In addition, it was noticed that the radial electric field could be produced by waves, generally through methods of interacting with trapped electrons that could not using various}

\section{Electric and magnetic rotational transforms}

\label{Sec:II}

Three sets of coordinates can be used to describe a toroidal magnetic field
and the associated particle orbits: ({\it i}) a cylindrical set of
coordinates around the toroidal $z$ vertical axis, $\left[ z,R,\varphi
\right] $ associated with the orthonormal basis $\left[ {\bf e}_{z},{\bf e}%
_{R},{\bf e}_{\varphi }\right] $; ({\it ii}) a local polar set of
coordinates around the magnetic axis $\left[ r,\theta \right] $ completed by
the toroidal angle $\varphi $ leading to $\left[ r,\theta ,\varphi \right] $
($R$ = $R_{0}+r\cos \theta $, $z$ = $r\sin \theta $) associated with the
orthonormal basis $\left[ {\bf e}_{r},{\bf e}_{\theta },{\bf e}_{\varphi
}\right] $; and ({\it iii}) a Cartesian set, $\left[ x,z\right] $ ($x$ = $%
r\cos \theta $, $z$ = $r\sin \theta $), of coordinates in the
radial/poloidal plane for each $\varphi $ with the local basis $\left[ {\bf e%
}_{x},{\bf e}_{z}\right] $. The toroidal magnetic field common to the
electric and magnetic rotational transform schemes is 
\begin{equation}
{\bf B_\phi}=\frac{B_{0}}{1+x/R_{0}}{\bf e}_{\varphi }\approx B_{0}{\bf e}%
_{\varphi }\text{.}  \label{bfbf}
\end{equation}
Finite aspect ratio effects associated with $x/R_{0}$ are analyzed in Appendix~\ref{Sec:appA}. 

\begin{figure}
\begin{center}
\includegraphics[width = 8cm]{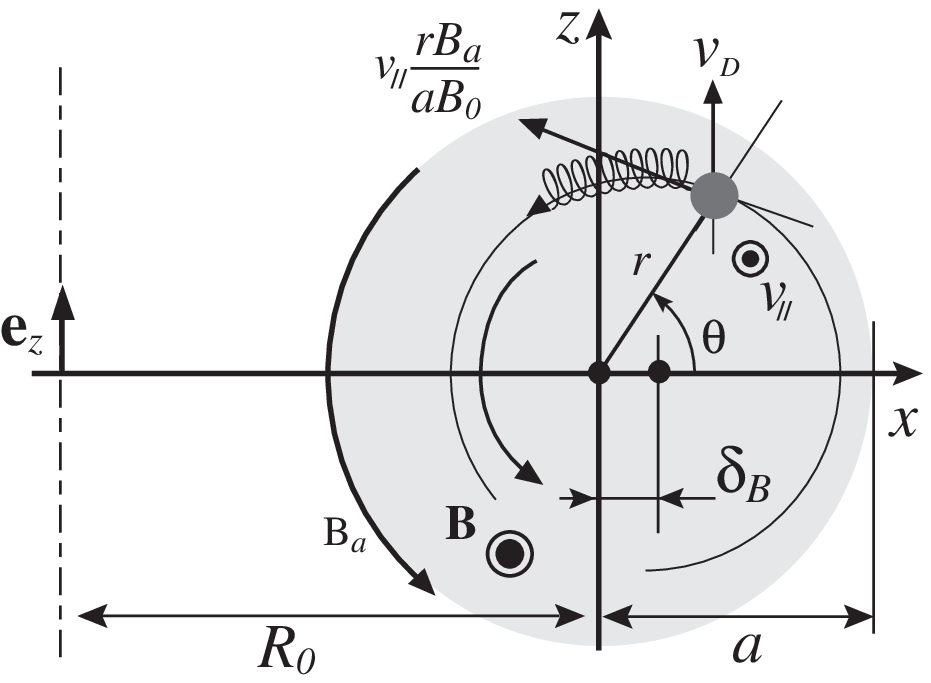}
\caption{Classical tokamak configuration with a poloidal magnetic field. }
\label{Fig:Figure2}
\end{center}
\end{figure}

Consider the simplest tokamak magnetic field
configuration displaying both toroidal, ${\bf B_\phi}$, and poloidal, 
\begin{equation}
{\bf B_p}=B_{a}\frac{r}{a}{\bf e}_{\theta }\text{,}
\end{equation}
magnetic components. 
Anticipating what might facilitate the comparison between electric
and magnetic rotational transforms, we  introduce $B_{a}$, the poloidal
field at the minor radius $r=a$, which we employ rather than the safety factor. 
This magnetic configuration is then fully characterized by two fields and two lengths $\left[
B_{0},B_{a},R_{0},a\right] $ and illustrated in Fig.~\ref{Fig:Figure2}. 
The velocity ${\bf v}$ of a charged particle in such a configuration is given by the 
sum of components along (${\bf v}_{\Vert }$), around (${\bf v}_{c}$) and
across (${\bf v}_{D}$) the field lines,
\begin{equation}
{\bf v}={\bf v}_{c}+{\bf v}_{\Vert }+{\bf v}_{D}\approx {\bf v}_{c}+v_{\Vert
}{\bf e}_{\varphi }+v_{\Vert }\frac{B_{a}}{B_{0}}\frac{r}{a}{\bf e}_{\theta
}+v_{D}{\bf e}_{z}.
\end{equation}
Dropping the fast cyclotron rotation ${\bf v}_{c}$, we end up with the
classical drift equations. These guiding center equations, restricted to the
poloidal plane $[x,z]$, can be expressed in term of the complex guiding
center variable ${\cal Z}$ in the drift equation: 
\begin{equation}
{\cal Z}=x+jz+a\frac{B_{0}}{B_{a}}\frac{v_{D}}{v_{\Vert }}\text{, }\frac{d%
{\cal Z}}{dt}=jv_{\Vert }\frac{B_{a}}{B_{0}}\frac{{\cal Z}}{a}.
\end{equation}

This drift equation is to be supplemented by the equation describing the diamagnetic
force along the field lines when the particle goes from the low field side toward the high field side, in the process converting the parallel linear momentum $mv_{\Vert }$ into cyclotron angular momentum $m{v_{c}}$.  %^{2}/\omega_{c}$ 
This diamagnetic conversion from linear momentum to cyclotron angular momentum
results in a low field side trapping if $\ v_{\Vert }/v_{c}<\sqrt{2r/R_{0}}$
in the equatorial plane. 
Here we have restricted the analysis to the passing
population in order to identify a single parameter aimed at comparing electric and
magnetic methods. 
The case of banana orbits is analyzed in Appendix~\ref{Sec:appA}.
That case generalizes the results obtained in this section.

The orbits of passing particles are circles, ${\cal Z}\left( t\right) $ = $%
{\cal Z}_{0}\exp (jv_{\Vert }B_{a}t/aB_{0})$.
The centers ($x=\delta_{B},z=0$) of the these drift orbits are shifted with respect to the
magnetic axis ($x=0,z=0$) by an amount $\delta _{B}$: 
\begin{equation}
\text{ }\frac{\delta _{B}}{a}=-\frac{v_{D}}{v_{\Vert }}\frac{B_{0}}{B_{a}}.
\label{db}
\end{equation}
These classical results, which can be found in any
standard textbook on tokamaks\cite{Rax2011}, now allow us to compare directly tokamak orbits with orbits in WDRT toroidal traps, where the poloidal magnetic field is replaced by a radial electric field.  

\begin{figure}
\begin{center}
\includegraphics[width = 8cm]{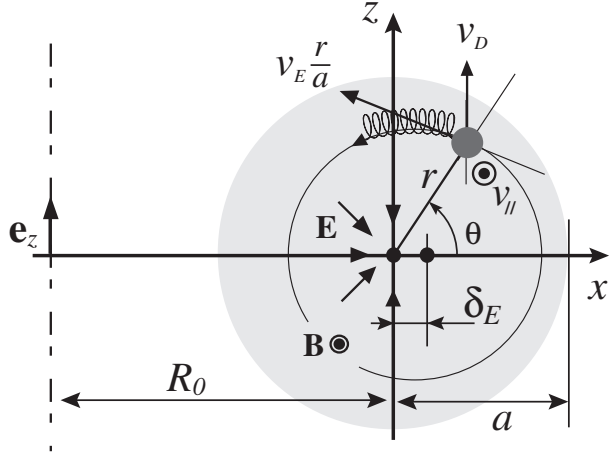}
\caption{Rigid body magneto-electric configuration with an electric rotational transform. }
\label{Fig:Figure3}
\end{center}
\end{figure}

Thus, consider the toroidal magnetic field ${\bf B_\phi}$, given in Eq.~(\ref{bfbf}), complemented by a radial electric field, 
\begin{equation}
{\bf E}=-E_{a}\frac{r}{a}{\bf e}_{r},
\end{equation}
where $-E_{a}$ is the value of the electric field at radius $r=a$. 
This configuration is fully characterized by two fields and two lengths $\left[
B_{0},E_{a},R_{0},a\right] $ and illustrated in Fig.~\ref{Fig:Figure3}. 
The velocity ${\bf v}$ of a charged particle in such a configuration is given by the 
sum of components along, around and across the field lines complemented by
the electric ${\bf E}\times {\bf B}$ drift, 
\begin{equation}
{\bf v}={\bf v}_{c}+{\bf v}_{\Vert }+{\bf v}_{D}+\frac{{\bf E}\times {\bf B}%
}{{B_{0}}^{2}}\approx {\bf v}_{c}+v_{\Vert }{\bf e}_{\varphi }+v_{D}{\bf e}%
_{z}+v_{E}\frac{r}{a}{\bf e}_{\theta },
\end{equation}
where we introduced $v_{E}${\it \ }$=E_{a}/B_{0}$ the electric drift
velocity at the outer edge of the plasma. The guiding center equations,
restricted to the poloidal $\left( x,z\right) $ plane can be analyzed with
the help of the complex guiding center variable ${\cal Z}$, which obeys
the drift equation: 
\begin{equation}
{\cal Z}=x+jz+a\frac{v_{D}}{v_{E}}\text{, }\frac{d{\cal Z}}{dt}=jv_{E}\frac{%
{\cal Z}}{a}.
\end{equation}
The solution, ${\cal Z}\left( t\right) $ = ${\cal Z}_{0}\exp (jv_{E}t/a)$,
shows that the orbits are circles whose centers ($x=\delta _{E},z=0$) are
shifted with respect to the magnetic axis ($x=0,z=0$) by an amount $\delta
_{E}$: 
\begin{equation}
\text{ }\frac{\delta _{E}}{a}=-\frac{v_{D}}{v_{E}}=-v_{D}\frac{B_{0}}{E_{a}}%
\approx \frac{a}{R_{0}}\frac{2k_{B}T}{eE_{a}a}.\text{{\it \ }}  \label{de}
\end{equation}
For a plasma temperature of the order of $\ k_{B}T/e\sim 10$ keV, in order
to achieve $\delta _{E}/a\ll 1$, the voltage drop between the center and the
edge of the discharge, $E_{a}a$, will be on the order of several hundred up
to perhaps one thousand kV.

The full analysis, presented in Appendix~\ref{Sec:appA}, confirms the simple results
obtained here. 
Consider a particle with initial parameters $\left( v_{\Vert_0},v_{c_0},x=x_{0},z=0\right)$:
 first ({\it i}) in a WDRT trap characterized by $\left[ B_{0},E_{a},R_{0},a\right]$;
  and then ({\it ii}) in a tokamak
configuration with parameters $\left[ B_{0},B_{a},R_{0},a\right] $.
According to the previous analysis and Appendix~\ref{Sec:appA}, the orbits are expected
to be similar provided that: 
$\delta _{E}\left(a,R_{0},E_{a},v_{\Vert_0},v_{c_0}\right) $ $\sim \delta _{B}\left(a,R_{0},B_{a},v_{\Vert_0},v_{c_0}\right) $. 
This can be viewed as a criteria that insures the same orbital confinement properties for the two configurations.
Thus let us consider ions, which display the largest magnetic shift $\delta _{B}$.
If we then average $v_{\Vert }$ over a thermal distribution ($\left\langle
{}\right\rangle $), we obtain the approximate similarity criteria, 
\begin{equation}
\left| \frac{\delta _{E}}{\delta _{B}}\right| \approx \left\langle \left|
v_{i\Vert }\right| \right\rangle \frac{B_{a}}{E_{a}}\sim 1.
\label{confinement}
\end{equation}
This simple criteria holds even for trapped particles.
As shown in Appendix~\ref{Sec:appA},   there are only passing particles and no
trapped banana orbits in the electric case. 
This is one of the main advantages of the electric drift compensation scheme. 
Note also that, since $E_{a}/B_{a}\sim \left\langle \left| v_{i\Vert }\right|
\right\rangle $, the edge rotation velocity $E_{a}/B_{0}$ is smaller than the
ion thermal velocity, with  $E_{a}/B_{0}\sim \left\langle \left| v_{i\Vert
}\right| \right\rangle $ $B_{a}/B_{0}$.

In the following, we consider hydrogen plasma, with electron and ion
densities $n_{e}$ and $n_{i}$, and masses $m_{e}$ and $m_{i}$. 
When there are no specification of the index $e$ or $i$, it means that the result stands for both species. 
With a radial voltage drop between the edge and the center of the order of one MV, such that $\delta _{E}/a\ll 1$, Gauss's theorem leads
to an estimate of the wave-driven radial space charge $e$($n_{i}$ $-$ $%
n_{e} $) $\sim $ $\varepsilon _{0}E_{a}/a$ $\sim $ $10^{-6}$ - $10^{-5}$ C/m$%
^{3}$, that is to say ($n_{i}$ $-$ $n_{e}$)/($n_{i}$ $+$ $n_{e}$) $\sim $ $%
10^{-7} $ - $10^{-6}$. 
This estimate is restricted to the body of the plasma. 
Note that the edge horizontal space charge, associated with the horizontal shift $\delta_{E}\neq 0$,  possibly leads to a horizontal electric field. 
This space  charge effect requires a somewhat separate consideration, given in Appendix~\ref{Sec:appD}.

%  \textcolor{red}{ PUT IN APPENDIX?}   This is presented in the next section.  

\section{Closed orbits with small displacement}
%\section{Inertial drift and alpha orbits}

\label{Sec:III}

We show here that it is possible  to satisfy simultaneously both closed orbits and small horizontal displacements.  
On the one hand, the radial electric field must be large enough to produce a rotational transform sufficient to overcome the vertical drift.
On the other hand, the electric field must not be quite that large so as to produce open orbits through inertial drifts.

In rotating plasma, both centrifugal forces and Coriolis forces lead to inertial drifts.
These drifts might lead to deconfinement above the so-called Brillouin limit~\cite{Brillouin1945,Davidson2001,Rax2015}. 
Consider a particle of charge $e$ and mass $m$,
interacting with ${\bf B=}B_{0}{\bf e}_{\varphi }$ and ${\bf E}=-E_{a}r{\bf e}_{r}/a$. 
Rather than the guiding center equations, we consider Newton's equations with both the electric and magnetic forces written with the complex variable ${\cal Z}$, 
\begin{equation}
{\cal Z}=x+jz,\text{ }\frac{d^{2}{\cal Z}}{dt^{2}}=-\frac{eE_{a}}{m}\frac{%
{\cal Z}}{a}-j\omega _{c}\frac{d{\cal Z}}{dt}\text{,}
\end{equation}
where, here, $x$ and $z$  no longer represent the guiding center position but rather represent the particle position. 
The solution to this classical problem is a superposition of the
slow and fast modes: ${\cal Z}$ = ${\cal Z}_{\pm }\exp j\omega _{\pm }t$,
where the slow and fast angular rotations velocities are given by: 
\begin{equation}
\frac{\omega _{\pm }}{\omega _{c}}=-\frac{1}{2}\mp \sqrt{\frac{1}{4}+\frac{%
eE_{a}}{ma{\omega _{c}}^{2}}}\text{.}  \label{id2}
\end{equation}
The fast rotation, 
\begin{equation}
\frac{\omega _{+}}{\omega _{c}}\approx -1+\frac{eE_{a}}{ma{\omega _{c}}^{2}}%
+...
\end{equation}
is not relevant here to analyze the drift motion~\cite{Brillouin1945,Davidson2001,Rax2015}. The slow one,
displaying also inertial drifts in addition to the $E_{a}r/aB_{0}$ rotation,
can be expanded according to: 
\begin{equation}
\frac{\omega _{-}}{\omega _{c}}\approx \frac{eE_{a}}{ma{\omega _{c}}^{2}}%
-\left( \frac{eE_{a}}{ma{\omega _{c}}^{2}}\right) ^{2}+...\text{.}  \label{id}
\end{equation}
We recognize the $E_{a}r/aB_{0}=$ $eE_{a}r/ma\omega _{c}$ as the so-called $E \times B$  rotation.   
The higher order term describes the inertial drifts~\cite{Rax2015}. 
Thus, to neglect the inertial drift, and validate the model of Sec.~\ref{Sec:II}, the electric field should
fullfil: $eE_{a}/ma\omega _{c}^{2}\ll 1$, or equivalently
$(eE_{a}a/kT)(\rho/a)^2 \ll 1$, where  $\rho _{i}$ is the ion Larmor radius. This provides an upper bound to $E_{a}$, which may be considered for solid body rotation equivalently as an upper bound to the voltage difference across the plasma.

But $E_{a}$ also satisfies the confinement condition from Eq.~$\left( \text{%
\ref{de}}\right) $, $\delta _{E}/a=2k_{B}T/eR_{0}E_{a}\ll 1$, which provides
a lower bound for $E_{a}$. 
Both upper and lower bounds must be compatible so that
\begin{equation}
\frac{2k_{B}T}{R_{0}}\ll eE_{a}\ll m_{i}a{\omega _{ci}}^{2}\text{.}  \label{lb}
\end{equation}
This relation requires the strong ordering $k_{B}T\ll m_{i}aR_{0}{\omega_{ci}}^{2}$. 
We can now  identify the constraint: $k_{B}T/m_{i}{\omega _{ci}}^{2}$ $\approx \rho _{i}^{2}$ $\ll R_{0}a$. 
Thus, there are at least six orders of magnitude between the upper
bound and the lower bound of Eq.~$\left( \text{\ref{lb}}\right) $. 
This very strong ordering allows one to chose easily an electric field $E_{a}$, large
enough to insure confinement ($\delta _{E}/a\ll 1$),  and yet small enough to
avoid inertial drifts ($eE_{a}/ma{\omega _{c}}^{2}\ll 1$), far below the
Brillouin limit. 
The two conditions, ({\it i}) confinement through the
requirement of closed orbits with small shift $\delta _{E}$ and ({\it ii})
negligible inertial drift effect, can thus be  simultaneously satisfied.

We address now the confinement and thermalization of energetic alpha particles in a WDRT trap. 
The considerations for energetic alpha particles are somewhat different than for plasma ions and electrons.
The simple analysis presented in Sec.~\ref{Sec:II} is no longer valid because: ({\it i}) we have to consider orbits whose starting
points are located near $x\sim z\sim 0$; and ({\it ii}) they start with a kinetic energy of $3.5$ MeV $\gg k_{B}T$ and slow down to the thermal energy
in perhaps a few hundreds of milliseconds. 
The equations describing the guiding center orbit taking into account energy slowing down are solved in Appendix~\ref{Sec:appB}.
The resulting $\left( x,z\right) $ poloidal orbit is given by: 
\begin{equation}
x(t)^{2}+z(t)^{2}=\frac{v_{D0}^{2}}{\nu _{\alpha}^{2}+(v_E/a)^2}\left[ 1-2\cos \left( \frac{v_{E}}{a%
}t\right) \exp \left( -\nu _{\alpha }t\right) +\exp \left( -2\nu _{\alpha
}t\right) \right],
\end{equation}
where $v_{D0}$ is the drift velocity at $t=0$ and $1/\nu _{\alpha }$ the
alpha particle slowing down time. 
An estimate of the various terms ($%
v_{E}>a\nu _{\alpha }$) then gives an upper bound to the radial extent
of the thermonuclear alpha particles: 
\begin{equation}
\text{Max }\left( \sqrt{x\left( t\right) ^{2}/a^{2}+z\left( t\right)
^{2}/a^{2}}\right) _{t}\approx \frac{a}{R_{0}}\frac{3.5\text{ MeV}}{eE_{a}a}%
<1.
\end{equation}

Thus, an edge to center voltage drop around one MV ensures thermal confinement and avoids inertial drifts.
The alpha particles will then be confined and thermalized in a WDRT. 
Note that the  heating of the alpha particles, mainly by collisions with electrons, will also be uniform on iso-potential surfaces.
Even if slowing down takes place preferentially on the low field side (or the high field side) due a horizontal displacement, the fast circular motion of thermal particles will ensure homogeneous energy deposition on the iso-potential surfaces.

\section{Energy content of radial and poloidal fields}

\label{Sec:IV}

We now compare the energy content of the WDRT to that of the equivalent tokamak.
Thus, consider a particle with charge $e$, mass $m$, and initial velocity and
position coordinates given by $\left( v_{\Vert_0},v_{c_0},x=x_{0},z=0\right) $, first in
a WDRT trap characterized by $\left[ B_{0},E_{a},R_{0},a\right] $ and then in a tokamak configuration with parameters $\left[
B_{0},B_{a},R_{0},a\right] $. 
According to the previous analysis, and generalized in further detail in Appendix~\ref{Sec:appA}, 
similar orbits will be described provided that  $\delta _{E}\left( a,R_{0},E_{a},{v_{\Vert_0}},{v_{c_0}}\right)
\sim $ $\delta _{B}\left( a,R_{0},B_{a},{v_{\Vert_0}},{v_{c_0}}\right) $, that is
to say, 
\begin{equation}
c^{2}\frac{\varepsilon _{0}E_{a}^{2}}{2}\sim \left\langle {v_{i\Vert
}}^{2}\right\rangle \frac{B_{a}^{2}}{2\mu _{0}}\text{.}  \label{cc}
\end{equation}
In writing Eq. $\left( \text{\ref{cc}}\right) $, we used Eqs. $\left( \text{%
\ref{db}}\right) $ and $\left( \text{\ref{de}}\right) $ and we took the ion
parallel velocity which corresponds to the largest magnetic shift $\delta
_{B} $.

Let us compare now, when Eq.~$\left( \text{\ref{cc}}\right) $ is satisfied, the energy content associated with the radial electric field $E_{a}$ and the poloidal magnetic field $B_{a}$. 
This energy content is a major source of free energy able to feed instabilities. 
While more detailed considerations are of course necessary, the energy content is important since we expect that, the lower the energy content, the more stable the configuration might tend to be and the least damaging might be any plasma disruption.

The energy content ratio can obtained on the basis of Eq.~$\left( \text{\ref{cc}}\right) $, 
%  but the electric term must be completed ({\it i}) with the rigid body rotation kinetic energy $Ma^{2}\Omega ^{2}/4$ ($\Omega =E_{a}/aB_{0}$ and $M$ is the mass of the plasma) and ({\it ii}) 
with the polarization energy associated with the radial permittivity of the plasma $\varepsilon _{\bot }$. 
The amount of energy needed to construct a radial electric field $E_{a}$ is then given by 
% the sum of the electrostatic energy, whose density is $\varepsilon _{\bot }\varepsilon _{0}E_{a}^{2}/2$, plus the kinetic energy,
%
\begin{equation}
\varepsilon _{\bot }\varepsilon _{0}\frac{{E_{a}}^{2}}{2}2\pi ^{2}R_{0}a^{2}=
\left( \frac{{\omega_{pi}}^{2}}{{\omega _{ci}}^{2}}\varepsilon _{0}\frac{{E_{a}}^{2}}{2}\right) 2\pi^{2}R_{0}a^{2}\text{.}
\end{equation}
Here we  used the low frequency ion permittivity, $\varepsilon
_{\bot }$ = ${\omega _{pi}}^{2}/{\omega _{ci}}^{2}$ ($\omega _{pi}$ and $\omega
_{ci}$ are the ion plasma and ion cyclotron frequencies). 
Note that the  low frequency ion permittivity captures both the electric energy content as well as the kinetic energy content through the rotation.
The ratio of the energy content to the magnetic energy in the equivalent tokamak content ${B_{a}}^{2}2\pi ^{2}R_{0}a^{2}/2\mu _{0}$, needed for the same orbital
confinement properties, is thus given by: 
\begin{equation}
\frac{{\omega _{pi}}^{2}}{{\omega _{ci}}^{2}}\frac{\varepsilon _{0}{E_{a}}^{2}}{%
\frac{{B_{a}}^{2}}{\mu _{0}}}\sim \frac{{\omega _{pi}}^{2}}{{\omega _{ci}}^{2}}%
\frac{\left\langle {v_{i\Vert }}^{2}\right\rangle }{c^{2}}\approx \left\langle
\beta \right\rangle <1,  \label{crite}
\end{equation}
where $\beta $ is the beta parameter of the tokamak, the ratio of thermal
energy to magnetic energy, which must be far smaller than one (typically less than 0.04 for a tokamak). This result
is very favorable to the electric option compared to the magnetic one.

Although we do not address directly  instabilities, or the transport of heat and particles through collisions, and particularly through turbulence,  a field configuration with less  free energy in the confining fields is promising.
It suggests that there may be less energy in any rearrangement of the magnetic surfaces in the plasma, together with their imbedded electric charge or toroidal current.
It suggests weaker turbulent activity, since the free energy drive for the instabilities, besides the thermal and particle gradients, is the free energy content of the  plasma. 
However, the comparative study of the main instabilities associated with a poloidal magnetic field configuration with respect to a radial
electric field configuration is  beyond the scope of this work.
Here we content ourselves with noting that the relatively  low free energy content is clearly promising in limiting the supply of energy to any instabilities associated with the wave-driven electric rotational transforms, while acknowledging that only a detailed analysis will ascertain whether this is in fact so.

\section{Wave-driven radial and toroidal particle dynamics}

\label{Sec:V}

The energy storage consideration is only part of the story; another key consideration is the power it takes to sustain the plasma fields.
Since the plasma is a dissipative media, both the radial electric field and the poloidal and magnetic field require power to persist.  
In order to maintain a poloidal magnetic field, charge must be transported along the magnetic field. 
In order to maintain a radial electric field, charge must be transported perpendicular to the magnetic field.
In either case, it is anticipated that  these fields may be maintained in the steady state by injecting waves into the plasma.

The transport of charge parallel to the magnetic field by rf waves is covered under the general theory of current drive by rf waves \cite{Fisch1987}.  In general, the current drive can be efficient through waves that provide toroidal momentum to the electrons or alter the collisionality of fast electrons \cite{fisch80}, which gives about the same efficiency.
The transport of charge perpendicular to the magnetic field can occur in a variety of ways.
It can happen by providing toroidal momentum to  electrons, which normally would support the toroidal current, but, if trapped, instead drift towards the magnetic axis \cite{fisch81a,Tsypin02}. 
The radial transport can also result from imparting perpendicular wave momentum to electrons \cite{Lee,guan13a,guan13b}, a mechanism that has been adduced to explain toroidal rotation by lower hybrid waves through generation of a radial electric field.

To compare the power dissipated in these cases, consider a simple plasma slab model with an homogeneous magnetic field
directed along the $z$ axis and an electromagnetic plane wave propagating in
the $\left( x,z\right) $ plane and interacting with a magnetized charged
particle with charge $e$ and mass $m$. The wave is a periodic function of
time and space described by the classical factor $\sin (k_{\bot }x+$ $k_{%
{\bf \Vert }}z-\omega t)$ where $\left( x,y,z\right) $ is a set of Cartesian
coordinates. Two regimes of interaction between this wave and a charged
particle are to be considered. ({\it i}) The adiabatic regime where the
energy/momentum exchange between the wave and the particle is small and
reversible. ({\it ii}) The resonant regime, when there exist an integer $N$
such that both the wave and the particle satisfy the relation: $\omega
=N\omega _{c}+k_{\Vert }v_{\Vert }$, where the exchanges are irreversible
and large. The regime of interest here is the resonant one.

When a resonant interaction takes place, energy, linear and angular momentum
are exchanged between the wave and the particle. 
This exchange results in a change of the energy $H$, the parallel momentum $mv_{\Vert }$ and the
cyclotron velocity $v_{c}$ of the particle, but also of the guiding center
position $y_{g}$, since part of the momentum is no longer free but bound to the
static magnetic field through the invariance of the canonical momentum along $y$.

This change in the guiding center position  provides orbital angular momentum deposition inside a magnetized plasma column and radial current generation. Appendix~\ref{Sec:appC} reviews the classical Hamilton equations
leading to the relations between energy, momentum and guiding center
dynamics, which can be summarized by the set~(\ref{Eq:C7}-\ref{Eq:C10}) of equations: 
\begin{eqnarray}
\frac{dy_{g}}{dH} &=&\frac{k_{\bot }}{m\omega _{c}\omega }\text{,}
\label{hh} \\
\frac{dv_{c}}{dH} &=&\frac{N\omega _{c}}{mv_{c}\omega }\text{,}  \label{hh4}
\\
\frac{dv_{\Vert }}{dH} &=&\frac{k_{\Vert }}{m\omega }\text{,}  \label{hh7}
\end{eqnarray}
where the energy variation $dH$ can be expressed as a function of $W$, the
steady state power transferred from the wave to the particle, simply as, $%
W=\partial H/\partial t$. 
This set of relations describing   wave-particle interactions was exploited in predicting the alpha channeling effect, where alpha particles are driven across field lines while cooled by waves \cite{Fisch1992,Fisch1994,Fetterman2008}.

\begin{figure}
\begin{center}
\includegraphics[width = 12cm]{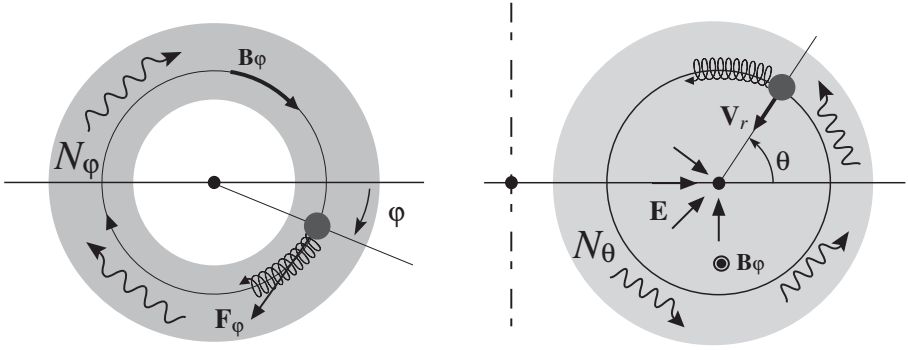}
\caption{Toroidal current generation ($N_{\varphi},F_{\varphi}$) and radial electric field generation ($N_{\theta},V_{r}$). }
\label{Fig:Figure4}
\end{center}
\end{figure}

To address toroidal and poloidal angular momentum exchanges in a  plasma torus,
rather than in a slab, we introduce the poloidal wave index $N_{\theta }$
and the toroidal wave index $N_{\varphi }$, 
\begin{eqnarray}
N_{\theta } &=&\frac{k_{\bot }c}{\omega }\text{, }  \label{vv} \\
N_{\varphi } &=&\frac{k_{\Vert }c}{\omega },  \label{vv2}
\end{eqnarray}
as illustrated in Fig.~\ref{Fig:Figure4}.
Consider a situation where, at some point in the torus, a particle absorbs
an amount of power $W$. This gives a radial particle velocity $V_{r}$ and a
toroidal momentum input described by the force $F_{\varphi }$,
\begin{eqnarray}
eV_{r} &=&e\frac{dy_{g}}{dt}=e\frac{dy_{g}}{dH}\frac{\partial H}{\partial t}=%
\frac{N_{\theta }}{cB_{0}}W_{E},  \label{dfgtr} \\
F_{\varphi } &=&m_{e}\frac{dv_{\Vert }}{dt}=m\frac{dv_{\Vert }}{dH}\frac{%
\partial H}{\partial t}=\frac{N_{\varphi }}{c}W_{B}\text{.}  \label{col2}
\end{eqnarray}
In deriving Eqs.~$\left( \text{\ref{dfgtr}}\right) $ and $\left( \text{\ref
{col2}}\right) $, we used Eqs.~$\left( \text{\ref{hh}-\ref{vv2}}\right) $ and
introduced $W_{E}$ (the absorbed wave power associated with radial electric
field generation in WDRT traps) and $W_{B}$ (the absorbed wave power
associated with poloidal magnetic field generation in classical tokamaks).
These results are not surprising;  in a magnetized plasma, momentum transfer
along the field lines results in a force and momentum transfer
perpendicular to the field lines results in a  velocity.

%\section{Efficiencies of wave-driven radial and toroidal current generation}
\section{Power consumption}
\label{Sec:VI}

In steady state, both the  radial velocity $V_{r}$ and toroidal force $F_{\varphi }$, driven by resonant waves, are balanced by collisions. 
The associated charge and momentum balances allows us to express 
$E_{a}$ and $B_{a}$ as functions of the total absorbed wave power, thus
defining the efficiencies of the generation processes. We will use Eq.~$\left( \text{\ref{confinement}}\right)$ to compare the power requirement for the sustainment of a radial electric field versus the sustainment of a toroidal current.
To find the steady state wave-driven effects, we  use the coefficients for collision frequencies and  ion viscosity $\eta _{i}$ as follows:
\begin{eqnarray}
\nu _{ei}\left( T\right) &=&\frac{\sqrt{2}ne^{4}\log \Lambda }{12\pi
^{3/2}\sqrt{m_{e}}\left( k_{B}T\right)^{3/2}{\varepsilon_{0}}^{2}}\text{, }%
\nu _{ii}=\sqrt{\frac{m_{e}}{m_{i}}}\nu _{ei}\text{, }  \label{nu1} \\
\text{ }\frac{\nu _{fi}\left( v\right) }{\nu _{ei}\left( T\right) } &=&\frac{%
3\sqrt{\pi }}{4}\left( \frac{2k_{B}T}{m_{e}v^{2}}\right) ^{3/2}\text{, }%
\frac{\eta _{i}}{\nu _{ii}}=\frac{3nk_{B}T}{10\sqrt{2}{\omega _{ci}}^{2}},
\label{nu2}
\end{eqnarray}
where $\nu _{ei}$ is the electron-ion momentum exchange frequency, $\nu
_{ii} $ the ion-ion momentum exchange frequency and $\nu _{fi}$ the fast
electron pitch angle scattering frequency at velocity $v$.

The $F_{\varphi }$ momentum input is balanced by electron-ion friction along the magnetic field lines.  This momentum balance provides the classical current drive efficiency~\cite{Fisch1978,Fisch1987}. 
The $F_{\varphi }$ wave drive, Eq.~(\ref{col2}), is balanced by the electron-ion friction
according to the relation:

\begin{equation}
\frac{N_{\varphi }}{c}W_{B}=m_{e}\nu _{fi}(c/N_{\varphi })v_{\Vert }\text{,}
\label{relb}
\end{equation}
where we took into account the fact that the driven electrons are resonant with the wave. 
The collision frequency is thus given by $\nu_{fi}(c/N_{\varphi })$. 
A suprathermal electron sustained with the velocity $v_{\Vert }$ along the toroidal direction drives a current $ev_{\Vert }/2\pi R_{0}$. Thus, 
\begin{equation}
2\pi aB_{a}=\mu _{0}\sum \frac{ev_{\Vert }}{2\pi R_{0}}=\frac{\mu _{0}e}{%
2\pi R_{0}}\frac{N_{\varphi }}{cm_{e}\nu _{fi}}\sum W_{B},  \label{relc}
\end{equation}
where we used Amp\`{e}re relation to establish the relation between the sum $%
\Sigma $ of all the currents and $B_{a}$. 
The  power needed to sustain  the poloidal field  is denoted by ${P_{B}}^{abs}$ =$\sum W_{B}$, where
\begin{equation}
{P_{B}}^{abs}\left( B_{a}\right) =\frac{4\pi ^{2}m_{e}}{e\mu _{0}}%
B_{a}aR_{0}\nu _{fi}(c/N_{\varphi })c/N_{\varphi }\text{.}  \label{pbbb}
\end{equation}
The proportionality between this power dissipated and the current sustained is the classical current efficiency~\cite{Fisch1978,Fisch1987}.

The power needed to sustain  the radial field is denoted by ${P_{E}}^{abs}=\Sigma W_{E}$. The dissipation mechanism for $W_{E}$ is less direct to calculate. We specialize to the limit of large aspect ratio, where the torus can be approximated as plasma column. In a rotating plasma column, absent centrifugal forces, electrons and ions would rotate at the same velocity. However, the difference in centrifugal forces leads to a differential in rotation velocities. The friction between electrons and ions then leads to power dissipation. This will give the minimum power dissipated;  additional dissipation can be expected in a finite aspect ratio torus.

In an infinite homogeneous fully ionized magnetized plasma, the collisions
between electrons and ions are unable to dissipate a steady electric field
perpendicular to the magnetic field because the $E$ cross $B$ drift is
ambipolar,  conservative ($\mathbf{E}\times \mathbf{B}\cdot \mathbf{B}=0$)
and does not provide the relative velocity between ions and electrons needed for friction
and dissipation.

On the other hand, in an inhomogeneous fully ionized magnetized plasma, perpendicular
conductivity is associated with the gradient and shear of the $%
E$ cross $B$ velocity. These gradient and shear arise from the inertia and viscosity terms
in the steady state sum of the ions momentum balance plus the electrons
momentum balance
\begin{equation}
m_{i}n_{i}\mathbf{v}_{i}\mathbf{\cdot }\nabla \mathbf{v}_{i}=\mathbf{j}%
\times \mathbf{B}-\mathbf{\nabla }\cdot \mathbf{\pi }\left( \mathbf{v}%
_{i}\right),
\label{Eq:34}
\end{equation}
where $\mathbf{\pi }$ is the viscous stress tensor, and a uniform density has been assumed. In this momentum balance, the pressure
force, which is responsible for the diamagnetic currents, has been neglected since no pressure dynamics is
driven by the electric field. Furthermore the radial component of the
current is insensitive to the radial gradient of the pressure. This equation
can be solved with respect to $\mathbf{j}$: 
\begin{equation}
\mathbf{j}=\frac{\mathbf{B}}{B^{2}}\times \mathbf{\nabla }\cdot 
\mathbf{\pi }\left( \mathbf{v}_{i}\right)+\frac{m_{i}n_{i}}{B^{2}}\mathbf{B}\times \mathbf{v}_{i}\mathbf{%
\cdot \nabla v}_{i}.
\end{equation}
If both $\nabla \mathbf{v}_{i}$ and $\mathbf{\pi }\left( \mathbf{v}%
_{i}\right) $ are driven by an inhomogeneous steady perpendicular electric
field, this relation is the nonlinear and nonlocal Ohm law in a fully
ionized magnetized plasma. The radial component $j_{r}$ involved in the
short-circuiting of the radial polarization is given by~\cite{Rozhansky2008,Helander2002}:
%Specifically, in a fully ionized collisional plasma column, rotating at velocity $v_{\theta }$, the
%expression of the dissipative radial current $j_{r}$ is given by~\cite{Rozhansky2008,Helander2002}, 
\begin{equation}
j_{r}=\frac{1}{B_{0}}\left( -\frac{\eta _{i}}{r}\frac{\partial }{\partial r}r%
\frac{\partial }{\partial r}v_{\theta }+m_{i}n\frac{v_{r}}{r}\frac{\partial 
}{\partial r}rv_{\theta }\right) \text{.}  \label{seh}
\end{equation}
The first term on the right hand side is associated with viscosity and the
second one with inertia.  In order to calculate $v_{r}$, consider the inertial effect described by Eqs.~$\left( \text{\ref{id2}}\right) $ and $\left( \text{\ref{id}}\right) $
responsible for a small difference between ion rotation and electron
rotation. This poloidal velocity difference $\left| v_{\theta i}-v_{\theta
e}\right| $ is given by: 
\begin{equation}
\left| v_{\theta i}-v_{\theta e}\right| =\left( \frac{eE_{a}}{a}\right)
^{2}\left| \frac{1}{{m_{i}}^{2}{\omega _{ci}}^{3}}-\frac{1}{{m_{e}}^{2}{\omega
_{ce}}^{3}}\right| r\approx \left( \frac{eE_{a}}{m_{i}a{\omega _{ci}}^{2}}%
\right) ^{2}\omega _{ci}r=\varepsilon ^{2}\omega _{ci}r\text{,}
\end{equation}
where we have defined the small parameter $\varepsilon =eE_{a}/m_{i}a{\omega
_{ci}}^{2}$ associated with inertial effects analyzed in Sec.~\ref{Sec:III}. Then,
electron-ion friction provides a poloidal force on both populations
(satisfying momentum conservation) which is responsible for a radial
ambipolar flow velocity $v_{r}$, 
\begin{equation}
\left| v_{r}\right| =\frac{m_{e}\nu _{ei}\left| v_{\theta i}-v_{\theta
e}\right| }{eB_{0}}=\varepsilon ^{2}\frac{m_{e}}{m_{i}}\nu _{ei}r,
\end{equation}
Both this slow radial ambipolar flow $v_{r}$ and the driven fast rigid-body
rotation $v_{\theta }$, 
\begin{equation}
\left| v_{\theta }\right| =\frac{E_{a}}{B_{0}}\frac{r}{a}=\varepsilon \omega
_{ci}r.
\end{equation}
are involved in the dissipation of the radial electric field through the
radial current described by Eq.~$\left( \text{\ref{seh}}\right) $. To
simplify the analysis, the temperature and density are assumed homogeneous.
Both currents on the right hand side of Eq.~$\left( \text{\ref{seh}}\right) $
are of the same order of magnitude and point in the same direction to short
circuit the wave-driven electric field.
\begin{eqnarray}
\left| j_{r}\right|  &=&\varepsilon \frac{nm_{i}\omega _{ci}\nu _{ei}}{B_{0}}%
\sqrt{\frac{m_{e}}{m_{i}}}\left( \frac{3}{10\sqrt{2}}\frac{{\rho _{i}}^{2}}{r}%
+2\varepsilon ^{2}\sqrt{\frac{m_{e}}{m_{i}}}r\right) \text{,} \\
\left\langle \left| j_{r}\right| \right\rangle _{a} &=&\varepsilon \frac{%
nm_{i}\omega _{ci}\nu _{ei}a}{B_{0}}\sqrt{\frac{m_{e}}{m_{i}}}\left( \frac{3%
}{5\sqrt{2}}\left( \frac{\rho _{i}}{a}\right) ^{2}+\frac{4}{3}\varepsilon ^{2}%
\sqrt{\frac{m_{e}}{m_{i}}}\right) \text{.}
\end{eqnarray}
We have introduced the ion Larmor radius ${\rho _{i}}=\sqrt{k_{B}T/m_{i}}/{\omega
_{ci}}$ and then averaged $\left\langle {}\right\rangle _{a}$ over the
plasma column, from the magnetic axis $r=0$ toward the edge $r=a$.  We
consider the wave drive expressed by Eq.~$\left(\text{\ref{dfgtr}}\right) $%
, and sum the power over the full discharge to balance $\left\langle \left|
j_{r}\right| \right\rangle _{a}$, 
\begin{equation}
{P_{E}}^{abs}\left( E_{a}\right) =2\pi ^{2}\varepsilon \frac{c}{N_{\theta }}%
nm_{i}\omega _{ci}\nu _{ei}a^{3}R_{0}\sqrt{\frac{m_{e}}{m_{i}}}\left( \frac{3%
}{5\sqrt{2}}\left( \frac{\rho _{i}}{a}\right) ^{2}+\frac{4}{3}\varepsilon ^{2}%
\sqrt{\frac{m_{e}}{m_{i}}}\right) \text{.}  \label{peee}
\end{equation}
The first term is associated with the viscous damping ($\mathbf{\nabla }%
\cdot \mathbf{\pi }$) of the electric polarization, while the second term results from the inertial effects ($\mathbf{v}_{i}\mathbf{\cdot \nabla v}_{i}$). Although the inertial term will be dropped in the conclusion, both terms are kept here because the small ratio $\left( \rho _{i}/a\right) ^{2}$ and $%
\varepsilon ^{2}\sqrt{m_{e}/m_{i}}$ can be of the same order of magnitude. This last result describes the wave-driven radial electric field generation
efficiency.

\section{Comparison of electric and magnetic rotational transforms}

\label{Sec:VII}

The two relations Eqs. $\left( \text{\ref{pbbb}}\right) $ and $\left( \text{%
\ref{peee}}\right) $ allow us to compare the two schemes on the basis of the
criteria Eq.~$\left( \text{\ref{confinement}}\right) $. We first obtain the
ratio, 
\begin{equation}
\frac{{P_{E}}^{abs}}{{P_{B}}^{abs}}=\varepsilon \frac{\nu _{ei}}{\nu _{fi}}\sqrt{%
\frac{m_{e}}{m_{i}}}\frac{N_{\varphi }}{N_{\theta }}\frac{B_{0}}{B_{a}}%
\left( \frac{a}{\lambda _{pe}}\right) ^{2}\left( \frac{3}{10\sqrt{2}}\left( 
\frac{\rho _{i}}{a}\right) ^{2}+\frac{2}{3}\varepsilon ^{2}\sqrt{\frac{m_{e}%
}{m_{i}}}\right),
\end{equation}
where $\lambda _{pe}=c/\omega _{p}$ is the inertial skin depth. It can be
further simplified on the basis of the value of the ratio of the collision
frequencies $\nu _{ei}/\nu _{fi}$ given by Eqs.~$\left( \text{\ref{nu1}}%
\right) $ and $\left( \text{\ref{nu2}}\right) $ and then $B_{a}$ can be
expressed as a function of $\varepsilon $ when the criteria Eq. $\left( 
\text{\ref{crite}}\right) $ is satisfied, $B_{a}/B_{0}$ $=$ $\varepsilon
a\omega _{ci}/\left\langle \left| v_{i\Vert }\right| \right\rangle \sim $ $%
\varepsilon a/\rho _{i}$. Note that we can also eliminate this ratio through
the classical tokamak ordering $B_{a}/B_{0}\sim a/2R_{0}$. The final result, 
\begin{equation}
\frac{{P_{E}}^{abs}\left( N_{\theta }\right) }{{P_{B}}^{abs}\left( N_{\varphi
}\right) }=\frac{4}{3N_{\theta }N_{\varphi }^{2}}\sqrt{\frac{m_{e}}{\pi m_{i}%
}}\frac{\rho _{i}a}{{\lambda _{pe}}^{2}}\left( \frac{m_{e}c^{2}}{2k_{B}T}%
\right) ^{3/2}\left( \frac{3}{10\sqrt{2}}\left( \frac{\rho _{i}}{a}\right)
^{2}+\frac{4}{6}\varepsilon ^{2}\sqrt{\frac{m_{e}}{m_{i}}}\right)\text{%
,}  \label{peee1}
\end{equation}
must be considered when $\varepsilon =E_{a}/B_{0}a\omega _{ci}$ takes its
typical values around $10^{-3}$. For this range of values, the double
ordering Eq.~$\left( \text{\ref{lb}}\right) $ is satisfied with typical
tokamak field values of the order of few Tesla and major and minor radius of
the order of few meters. This relation gives the ratio of wave power needed
to provide the same orbital confinement with two configurations first ({\it i%
}) in a WDRT trap characterized by $\left[ B_{0},E_{a},R_{0},a\right] $ and
then ({\it ii}) in a tokamak configuration with parameters $\left[
B_{0},B_{a},R_{0},a\right] $, the electric field appear in the small
parameter $\varepsilon =E_{a}/B_{0}a\omega _{ci}$ and the magnetic poloidal
field has been eliminated with the similarity criteria. Keeping only the
dominant basic scaling we can write: 
\begin{equation}
\frac{{P_{E}}^{abs}}{{P_{B}}^{abs}}\sim \frac{1}{N_{\theta }N_{\varphi }^{2}}%
\left( \frac{a}{\lambda _{pe}}\right) ^{2}\left( \frac{\rho _{i}}{a}\right)
^{3}\left( \frac{m_{e}c^{2}}{k_{B}T}\right) ^{3/2}\sqrt{\frac{m_{e}}{m_{i}}}%
\text{.}
\end{equation}
This last scaling can be evaluated on the basis of the typical values: $%
N_{\theta }^{-1}N_{\varphi }^{2}\sim 10^{-1}$, $a/\lambda _{pe}\sim 10^{3}$, 
$\rho _{i}/a\sim 10^{-3}$, $mc^{2}/k_{B}T\sim 10^{2}$, thus the expected
favorable ordering $P_{E}^{abs}<P_{B}^{abs}$ is confirmed. 

%This ordering ${P_{E}}^{abs}<{P_{B}}^{abs}$ is favorable to the electric scheme, particularly near plasma cold resonances where $N_{\theta }$ becomes very large.  \textcolor{red}{ are we sure that it works like that? it may be that only the momentum coming from the waveguide counts.  }
%
%Even far from these resonances, the small effective conductivity perpendicular to the magnetic field is responsible for the small power
%needed to cancel the viscous and inertial currents Eq. $\left( \text{\ref {seh}}\right) $ short circuiting the wave-driven radial electric field.
%\textcolor{red}{I dont understand this point. probably you need to explain what you mean by short circuit.}

\section{Summary and conclusions}
\label{Sec:Summary}

The main  result of this study is the suggestion that a wave-driven radial electric field to produce the rotational transform requires less energy storage and less power dissipation (at large aspect ratios) than the wave-driven toroidal electric current that produces the rotational transform in a conventional, but steady state, tokamak.   These results are
described by Eqs.~$\left( \text{\ref{crite}, \ref{peee} and \ref{peee1}}%
\right) $.
 
However, our level of understanding of WRDT properties is far below our present level of understanding of tokamak properties. Fluid and kinetic models have  been widely used to turn the tokamak concept into an operational machine. The encouraging results presented in this speculative analysis at most indicate the necessity to follow the same lines of kinetic and fluid modeling to assess the full potential of WRDT for thermonuclear fusion.

A number of issues clearly need further study. 
The pressure balance in a WDRT needs to be constructed; this balance will likely require at least a small toroidal current together with a vertical magnetic field.
The instabilities of a WDRT configuration need to be identified.
Methods to stabilize the main large scale instabilities need to be identified as well. 
The confinement time associated with the remaining level of turbulent instabilities needs to be evaluated. 
The additional dissipation due to finite aspect ratio effects needs to be evaluated.

Particularly in view of the additional dissipation expected in the case of finite aspect ratio, it remains to be explored how might the radial electric field profile be optimized. 
Here we considered  rigid-body rotation, $\partial \Omega /\partial r=0$ and $v_{\theta }\sim \Omega r$, where electron-ion poloidal friction is minimized to second order in $\varepsilon $. 
This poloidal velocity field displays a local velocity shear leading to viscous effects. 
It may be that larger perhaps less viscous damping can be obtained, possibly at the expense of increased electron-ion
poloidal friction, using other rotation profiles, for example, like the
Keplerian rotation, $\Omega \sim 1/r$ and $\partial v_{\theta }/\partial r=0$.  
There is opportunity, in principle, to chose this rotation profile through the wave damping profile.
%It remains to be explored how the poloidal rotation is accelerated on the high field side and decelerated on the low field side to comply with the toroidal geometry. 
%Besides this probable finite aspect ratio drawback, a
%direction to improve this original scheme is given by the possibility to
%optimize the electric field profile, that is to say the rotation profile $%
%\Omega \left( r\right) $ which is associated with the wave power deposition
%profile $W_{E}\left( r\right) $. Here we have explored the rigid-body
%rotation, $\partial \Omega /\partial r=0$ and $v_{\theta }\sim \Omega r$,
%where electron-ion poloidal friction is minimized to second order in $%
%\varepsilon $. 
%
%But the poloidal velocity field display a local velocity
%shear leading to viscous effect. Besides rigid-body, the other common type
%of rotation in fluid dynamics is Keplerian rotation, $\Omega \sim 1/r$ and $%
%\partial v_{\theta }/\partial r=0$, which imply a larger electron-ion
%poloidal friction but less viscous damping. The two terms on the right hand
%side of Eq. $\left( \text{\ref{seh}}\right) $ are here of the same order of
%magnitude, but $W_{E}\left( r\right) $ can be tuned to preferentially
%increase or decrease one of these terms, possibly leading to power
%dissipation minimization. Such an optimization requires to take into account
%all the density and temperature profile.
%
%
In that respect, in addition to the other issues requiring further study, what remains to be identified are also the waves that produce the radial transport of charge, together with their propagation characteristics in rotating media.
In this respect, we expect to find useful waves that were predicted to be useful for alpha channeling, such as the ion Bernstein wave  \cite{fisch95b},  a combination of waves  \cite{fisch95,herrmann97}, or, considering the rotating geometry, even stationary waves \cite{fetterman_stationary}.

It may be that the real potential of the two very favorable orderings, namely  Eqs.~$\left( \text{\ref{crite}}\right) $ and $\left( \text{\ref{peee1}}\right) $, will turn out to be most useful within a hybrid steady state configuration, where part of the rotational transform is achieved with a wave driven toroidal current at the edge, and the other part with a wave driven radial electric field near the
center. In any event, what is hoped for here is that our identification of the possibility of achieving single particle confinement in a steady state toroidal trap, with low free energy and with low power dissipation, should be stimulatory for further consideration of all the outstanding issues in a wave-driven rotating torus.

\section*{Acknowledgement}

This work was supported by US DOE Contract No. DE-SC0016072.

\appendix

\section{Orbit invariants}

\label{Sec:appA}

Consider two similar toroidal fields, with the same magnetic field $B_{0}$
on the magnetic axis $R$ = $R_{0}$, ({\it i}) one is complemented with a
radial electric field with value $E_{a}$ at the edge, and ({\it ii}) one is
complemented with a poloidal magnetic filed with value $B_{a}$ at the edge,
both fields being linear with respect to the radius $r$.

Then, consider a charged particle of mass $m$ and charge $e$ with velocities 
${v_{\Vert_0}}$ and ${v_{c_0}}$ at the point ($x=x_{0}$, $z=0$) on the low field
side, first ({\it i}) in a WDRT trap characterized by $\left[
B_{0},E_{a},R_{0},a\right] $ and then ({\it ii}) in a tokamak configuration
with parameters $\left[ B_{0},B_{a},R_{0},a\right] $. We will address here
the issue of the comparisons of orbits in both confinement schemes in its
full generality and thus use the orbit invariants in order to take into
account the toroidal magnetic field inhomogeneity $B_{\varphi }$ = $%
B_{0}/\left( 1+x/R_{0}\right) $. We will show that the two fields-particles
interactions problems are completely characterized by the parameters $\delta
_{E}\left( a,R_{0},E_{a},{v_{\Vert_0}},{v_{c_0}}\right) $ and $\delta _{B}\left(
a,R_{0},B_{a},{v_{\Vert_0}},{v_{c_0}}\right) $ so that to compare the orbital
confinement properties of both configurations we have to state simply the
criteria $\delta _{E}=\delta _{B}$. For the very same initial position, and
initial parallel and cyclotron velocities, a given particle will display
almost the same orbit within the two configuration provides that $\delta
_{E}=\delta _{B}$, this simple criteria will be confirmed even for trapped
particles as the previous analysis in Sec.~\ref{Sec:II} was restricted to passing
particles and neglected the fact that the parallel velocity is modulated by
the diamagnetic force during its transit from the low field side ($x>0$)
toward the high field side ($x<0$). If this modulation reaches the zero
parallel velocity level, for the magnetic case we have to consider banana
orbits on the low field side and potatoes orbits near the center, and for
the electric case we will show that the conservation of energy does not
imply a cancelation of the parallel velocity as the electrostatic energy
comes into play to insure energy conservation. The parameters $\delta
_{E}\left( a,R_{0},E_{a},{v_{\Vert_0}},{v_{c_0}}\right) $ and $\delta _{B}\left(
a,R_{0},B_{a},{v_{\Vert_0}},{v_{c_0}}\right) $ are defined as positive
quantities: 
\begin{equation}
\frac{\delta _{E}}{a}=m\frac{{v_{\Vert_0}}^{2}+{v_{c_0}}^{2}/2}{R_{0}\left|
e\right| E_{a}}\text{, }\frac{\delta _{B}}{a}=m\frac{{v_{\Vert_0}}^{2}+{v_{c_0}}^{2}/2}{R_{0}\left| e{v_{\Vert_0}}\right| B_{a}}\text{.}
\end{equation}
The initial pitch angle variable $0\leq \chi _{0}\left( v_{\Vert_0},{v_{c_0}}\right) \leq 2$ is defined according to the ratio: 
\begin{equation}
\chi _{0}=\frac{{v_{c_0}}^{2}}{{v_{\Vert_0}}^{2}+{v_{c_0}}^{2}/2}\text{.}
\end{equation}
The identification of a similarity criteria between electric and magnetic
rotational transforms can be achieved through the analysis of the radial
extent of the orbits in the poloidal/radial plane when $\chi _{0}$ is varied
within the range $\left( 0,2\right) $. In order to further simplify the
analysis we will consider $R_{0}$ as the unit of length in the remaining
part of this Appendix.

Three invariants can be identified in a static toroidal field configuration:
({\it i}) the magnetic moment, ({\it ii}) the toroidal angular momentum and (%
{\it iii}) the particle energy. The magnetic moment is conserved under
adiabatic hypothesis, all along the trajectory for both the magnetic and the
WDRT fields configurations, this invariance, ${v_{c}}^{2}R$ $=$ $%
{v_{c_0}}^{2}R_{|_{t=0}}$, can be express through the relation Eq. $\left( \text{\ref
{mom}}\right) $, 
\begin{equation}
\frac{{v_{c}}^{2}}{{v_{c_0}}^{2}}=\frac{1+x_{0}}{1+x}\text{.}  \label{mom}
\end{equation}

The toroidal angular momentum is also an invariant, but for the magnetic
case we have to add up the toroidal component of the vector potential to the
kinetic angular momentum, that is to say add up the toroidal flux of the
poloidal field. So, we have to consider the relation $v_{\Vert }R$ = $%
{v_{\Vert_0}}R_{|_{t=0}}$ for the electric case and add the poloidal field flux for
the magnetic ones, this leads to the relations:

\begin{eqnarray}
\frac{v_{\Vert }}{{v_{\Vert_0}}} &=&\frac{1+x_{0}}{1+x}\text{,}  \label{ele} \\
\frac{v_{\Vert }}{{v_{\Vert_0}}} &=&\frac{1+x_{0}}{1+x}\pm \frac{1}{2-\chi _{0}%
}\frac{1}{1+x}\frac{z^{2}+x^{2}-{x_{0}}^{2}}{\delta _{B}}\text{,}  \label{ma}
\end{eqnarray}
where the first one Eq. $\left( \text{\ref{ele}}\right) $ is associated with
WDRT traps and the second one Eq. $\left( \text{\ref{ma}}\right) $ with
tokamak configurations, note that $R_{0}$ is the unit of length. Energy is
also conserved in static fields, kinetic energy for tokamaks, Eq. $\left( 
\text{\ref{mag}}\right) $, and kinetic energy plus electrostatic potential
energy, Eq. $\left( \text{\ref{elec}}\right) $, for the WDRT case:

\begin{eqnarray}
{v_{c}}^{2}+{v_{\Vert }}^{2} &=&{v_{c_0}}^{2}+{v_{\Vert_0}}^{2}\left( 1\pm \frac{2}{%
2-\chi _{0}}\frac{z^{2}+x^{2}-{x_{0}}^{2}}{\delta _{E}}\right) \text{,}
\label{elec} \\
{v_{c}}^{2}+{v_{\Vert }}^{2} &=&{v_{c_0}}^{2}+{v_{\Vert_0}}^{2}\text{.}  \label{mag}
\end{eqnarray}
For a given particle $\left( x_{0},\chi _{0},\delta \right) $, we have four
dynamical variables $\left( v_{\Vert },v_{c}\right) $ and $\left( x,z\right)
,$ and three relations, Eqs. $\left( \text{\ref{mom}}\right) $, $\left( 
\text{\ref{ele}}\right) $ and $\left( \text{\ref{elec}}\right) $ or Eqs. $%
\left( \text{\ref{mom}}\right) $, $\left( \text{\ref{ma}}\right) $ and $%
\left( \text{\ref{mag}}\right) $, between these variables, thus we can
eliminate the velocity $\left( v_{\Vert },v_{c}\right) $ and find the
Cartesian equations describing the orbit in the $\left( x,z\right) $ plane
for this particle with initial and structural set of conditions $\left(
x_{0},\chi _{0},\delta _{E}\right) $ or $\left( x_{0},\chi _{0},\delta
_{B}\right) $: 
\begin{eqnarray}
\pm \frac{z^{2}+x^{2}-{x_{0}}^{2}}{\delta _{E}}\left( 1+x\right) ^{2} &=&\chi
_{0}\left( x_{0}-x\right) \left( 1+x\right) \nonumber\\ & & \quad+\frac{2-\chi _{0}}{2}\left(
x_{0}-x\right) \left( 2+x_{0}+x\right) \text{,}  \label{b} \\
\left[ \pm \frac{z^{2}+x^{2}-{x_{0}}^{2}}{\delta _{B}}+\left( 2-\chi
_{0}\right) \left( 1+x_{0}\right) \right] ^{2} &=&2\chi _{0}\left( \chi
_{0}-2\right) \left( x_{0}-x\right) \left( 1+x\right) \nonumber\\ & & \quad+\left( 2-\chi
_{0}\right) ^{2}\left( 1+x\right) ^{2}\text{.}  \label{d}
\end{eqnarray}
The $\pm $ sign is associated with the sign of the product $e{v_{\Vert_0}}$
for the tokamak configuration and with the sign of the charge $e$ for the
WDRT configuration.

These two families of ovals are determined by the initial conditions $\left(
x_{0},\chi _{0}\right) $ and by a single structural parameter, $\delta _{E}$
or $\delta _{B}$. The equations of these two families of ovals allows to
compare the orbits of the two confinement schemes very easily by varying the
pitch angle $\chi _{0}$ and the initial position $x_{0}$ for a given ratio $%
\delta _{E}/\delta _{B}$. For example, for a copassing positively charged
particle, if $\delta _{E}/\delta _{B}$ is set equal to one and if $\chi _{0}$
is small, the orbits are similar as they can be approached by the relations
: $\left( z^{2}+x^{2}-{x_{0}}^{2}\right) /\delta _{E}$ $\approx $ $2\left(
x-x_{0}\right) $ and $\left( z^{2}+x^{2}-{x_{0}}^{2}\right) /\delta _{B}$ + $%
2\left( 1+x_{0}\right) $ $\approx $ $2\left( 1+x\right) $, where we have
used the ordering $\chi _{0}<1$, $x<1$ and $x_{0}<1$ as $R_{0}$ is the unit
of length. We recognize here the results of the previous drift theory,
obtained in Sec.~\ref{Sec:II}, up to a small ${\delta _{E}}^{2}$ term for the electric
case due to the toroidicity. For tokamak, we get two orbits for the same $%
\left( x_{0},{v_{\Vert_0}},{v_{c_0}}\right) $, or equivalently $\left( x_{0},\chi
_{0},\delta _{B}\right) $, set of parameters. As expected, these two ovals
will degenerate into a banana when merging together for large $\chi _{0}$,
one of them is similar to the electric orbit and the other display a larger
radial extension in the poloidal/radial plane. A similar asymptotic analysis
can be considered for $\chi _{0}\sim 2$ and we will end up with the
classical width of the banana and the fact that they can be contained into
the corresponding circular electric orbit or be outside it, depending on the
sign of the initial parallel velocity ${v_{\Vert_0}}$.

\begin{figure}
\begin{center}
\includegraphics[width = 9cm]{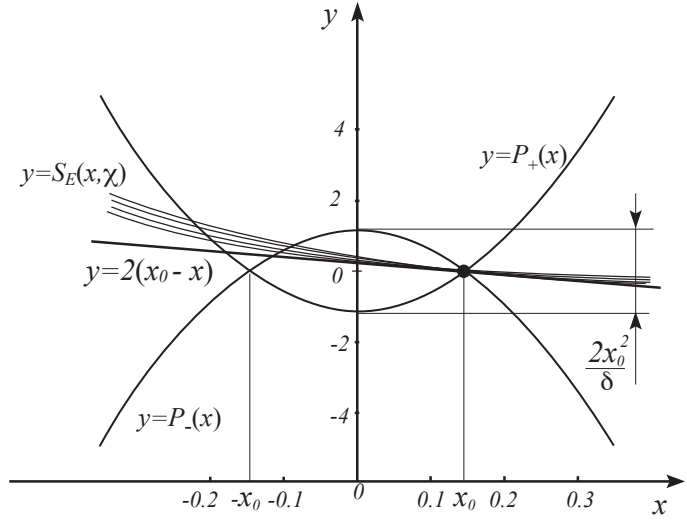}
\caption{Poloidal extension of particle orbits for magneto-electric confinement. }
\label{Fig:Figure5}
\end{center}
\end{figure}

\begin{figure}
\begin{center}
\includegraphics[width = 9cm]{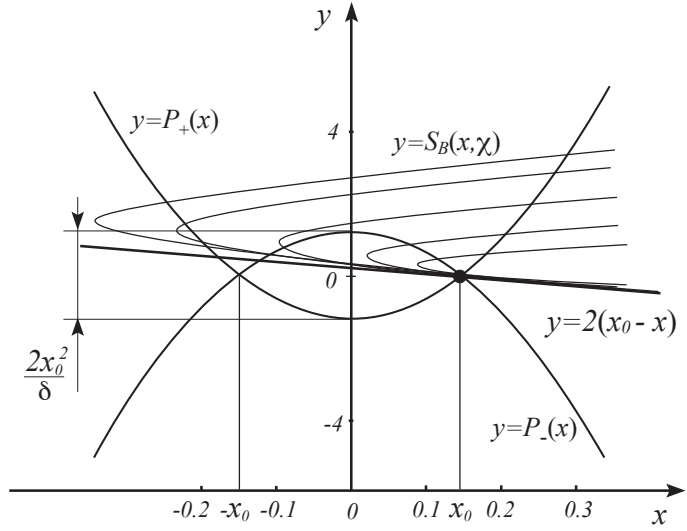}
\caption{Poloidal extension of particle orbits for tokamak confinement. }
\label{Fig:Figure6}
\end{center}
\end{figure}

The most straightforward and convincing analysis of orbital confinement can
be carried out on the basis of the equatorial radial size which can be
obtained through the requirement $z=0$ in the ovals equations Eq. $\left( 
\text{\ref{b}}\right) $ and Eq. $\left( \text{\ref{d}}\right) $. Thus, we
define two parabolic branches $P_{\pm }\left( x\right) $ as:

\begin{equation}
P_{\pm }\left( x\right) =\pm \frac{x^{2}-{x_{0}}^{2}}{\delta }\text{,}
\end{equation}
and two families of electric $S_{E}\left( x,\chi _{0}\right) $ and magnetic $%
S_{B}\left( x,\chi _{0}\right) $ functions restricted to $0\leq \chi
_{0}\leq 2$ :

\begin{eqnarray}
S_{E}\left( x,\chi _{0}\right) &=&\chi _{0}\left( x_{0}-x\right) \left(
1+x\right) +\frac{2-\chi _{0}}{2}\left( x_{0}-x\right) \left(
2+x_{0}+x\right) \text{,} \\
S_{B}\left( x,\chi _{0}\right) &=&\pm \left( 2-\chi _{0}\right) \left(
1+x_{0}\right) \nonumber\\ & & \quad\pm \sqrt{2\chi _{0}\left( \chi -2\right) \left(
x_{0}-x\right) \left( 1+x\right) +\left( 2-\chi _{0}\right) ^{2}\left(
1+x\right) ^{2}}\text{.}
\end{eqnarray}
The radial extent of an orbit in the equatorial plane is given by the
solution of the equations : ({\it i}) $P_{\pm }\left( x\right)
=S_{E}\left( x,\chi _{0}\right) $ in the electric case, and ({\it ii}) $%
P_{\pm }\left( x\right) =S_{B}\left( x,\chi _{0}\right)$ in the magnetic
one. A simple numerical scan over the values $0<\chi _{0}<2$ and $0<x_{0}\ll
1$ allows to explore the range of values of the parameter $\delta
_{E}/\delta _{B}$ needed to achieve the orbit similarity condition. Figures~\ref{Fig:Figure5} and \ref{Fig:Figure6} display the typical set of result obtained through this simple two
parameters numerical scan.

The specific numerical values associated with Fig.~\ref{Fig:Figure5} and Fig.~\ref{Fig:Figure6} are $%
x_{0}=0.15\times R_{0}$ and $\delta _{E}=\delta _{B}=0.02\times R_{0}$, the $%
\chi $ scan is $\left( 0.,0.5,1.,1.5,2\right) $ in Fig.~\ref{Fig:Figure5} and $\left(
0.8,1.,1.3,1.6,1.8\right) $ in Fig.~\ref{Fig:Figure6}. Although passing particles have the
same orbit provided that $\delta _{E}/\delta _{B}=1$, large banana orbits
and potatoes orbits have a larger radial extents under this condition, so
they require a smaller $\delta _{B}$ to fit with electric orbits with the
same $\left( {v_{\Vert_0}},{v_{c_0}},x_{0}\right) $. A global conclusion of the
numerical survey of $P_{\pm }=S_{B}$ can be stated as follows : {\it when a
given }$\left( {v_{\Vert_0}},{v_{c_0}},x_{0},z=0\right) ${\it \ particle is
confined in a WDRT configuration }$\left[ B_{0},R_{0},\delta _{E}\right] $%
{\it , the very same particle }$\left( {v_{\Vert_0}},{v_{c_0}},x_{0},z=0\right) $%
{\it \ is also confined in a tokamak configuration }$\left[
B_{0},R_{0},\delta _{B}\right] ${\it .}

\section{Alpha particles orbits}

\label{Sec:appB}

Consider a thermonuclear alpha particle ($\alpha $), with charge $2e$,
confined in a WDRT trap characterized by $\left[ B_{0},E_{a},R_{0},a\right] $%
. In this WDRT configuration, the velocity ${\bf v}$ of this $\alpha $
particle is given by the classical sum of components along (${\bf v}_{\Vert
} $), around (${\bf v}_{c}$) and across (${\bf v}_{D}$) the field lines
complemented by the $E/B$ rotation around the magnetic axis, 
\begin{equation}
{\bf v}={\bf v}_{c}+{\bf v}_{\Vert }+{\bf v}_{D}+\frac{{\bf E}\times {\bf B}%
}{B_{0}^{2}}\approx {\bf v}_{c}+v_{\Vert }{\bf e}_{\varphi }+v_{D}{\bf e}%
_{z}+v_{E}\frac{r}{a}{\bf e}_{\theta }.
\end{equation}
The initial velocity of this $\alpha $ particle is $c/23$ and in between $%
3.5 $ MeV and $0.5$ MeV it slows down on thermal electrons, in typically $%
0.1 $-$0.2$ second, according to the slowing down equation: 
\begin{equation}
\frac{d\left( {\bf v}_{c}+{\bf v}_{\Vert }\right) }{dt}=-\frac{\nu _{\alpha }%
}{2}\left( {\bf v}_{c}+{\bf v}_{\Vert }\right) ,  \label{relax}
\end{equation}
where the {\it energy relaxation frequency} $\nu _{\alpha }$ is given by the
relation~\cite{Rax2011}: 
\begin{equation}
\nu _{\alpha }=\frac{2\sqrt{2}\ln \Lambda }{3\pi ^{\frac{3}{2}}}\frac{%
Z_{i}e^{4}n_{i}}{\varepsilon _{0}^{2}m_{\alpha }m_{i}}\sqrt{\frac{m_{e}}{kT}}%
\frac{m_{i}}{kT},
\end{equation}
the notation are standard and $\ln \Lambda $ is the Coulomb logarithm.
Starting from $\left| {\bf v}_{c}+{\bf v}_{\Vert }\right| =c/23$, this
linear behavior of $\alpha $ relaxation in between $3.5$ MeV and $0.5$ MeV
allows to express the evolution of the drift velocity $v_{D}\left( t\right) $
as a function of the initial drift velocity $v_{D_0}=v_{D}\left( t=0\right) $%
, 
\begin{equation}
v_{D}\left( t\right) =\frac{{v_{\Vert }}^{2}\left( t\right) +{v_{c}}^{2}\left(t\right) /2}{R_{0}\omega _{c_0}}=v_{D_0}\exp -\nu _{\alpha }t\text{.}
\end{equation}
Then, below $500$ keV, thermal ions come into play and, in less than $0.1$
second, the ultimate phase of $\alpha $ thermalization from few hundreds keV
down to few tens of keV is achieved. It is to be noted that this final stage
of relaxation can be easily included in Eq. $\left( \text{\ref{relax}}%
\right) $ and $v_{D}\left( t\right) $ can be expressed as a function of $%
v_{D_0}$ and time, but this ultimate phase of thermalization does change the
physical picture of $\alpha $ confinement in WDRT. Following Sec.~\ref{Sec:II}, the
guiding center equations, restricted to the poloidal plane, can be analyzed
with the help of the complex guiding center variable ${\cal Z}=x+jz$ which
satisfies the drift/slowing down equation: 
\begin{equation}
\frac{d{\cal Z}}{dt}=j\omega _{E}{\cal Z}+jv_{D_0}\exp -\nu _{\alpha }t,
\label{relaxdrift}
\end{equation}
where we have defined $\omega _{E}=v_{E}/a$. Because the reactivity of $D$
and $T$ is a very steep function of density and temperature, most of the
fusion reactions takes place near $x\sim z\sim 0$. So, in order to simplify
the analysis, we consider the orbits starting at $t=0$ with the energy and
position initial conditions: $3.5$ MeV and ${\cal Z}=0$. The solution of Eq. 
$\left( \text{\ref{relaxdrift}}\right) $ with these initial conditions is: 
\begin{equation}
\text{ }{\cal Z}=jv_{D_0}\exp j\omega _{E}t\int_{0}^{t}du\exp -\left( \nu
_{\alpha }+j\omega _{E}\right) u=v_{D_0}\frac{\exp \left( -\nu _{\alpha
}t\right) -\exp \left( j\omega _{E}t\right) }{j\nu _{\alpha }-\omega _{E}},%
\text{{\it \ }}
\end{equation}
note that others initial condition just add up a rotating term in front of
the previous integral. The radial extension of this alpha particle orbit is
given by $x^{2}+z^{2}$ $=$ ${\cal ZZ}^{*}$: 
\begin{equation}
{\cal ZZ}^{*}=\frac{{v_{D_0}}^{2}}{\nu _{\alpha }^{2}+\omega _{E}^{2}}\left[
\exp \left( -2\nu _{\alpha }t\right) -2\cos \left( \omega _{E}t\right) \exp
\left( -\nu _{\alpha }t\right) +1\right]
\end{equation}
This radial excursion in the poloidal plane takes remains between the upper
and lower bounds: 
\begin{equation}
\frac{v_{D_0}}{\omega _{E}}\left( 1-\exp -\nu _{\alpha }t\right) \leq \sqrt{%
x^{2}+z^{2}}\leq \frac{v_{D_0}}{\omega _{E}}\left( 1+\exp -\nu _{\alpha
}t\right)
\end{equation}
where we have taken into account the ordering $\omega _{E}>\nu _{\alpha }$.
Starting from the lower bound at $t=0$, the maximum radial extension during
the thermalization process is of the order of:

\begin{equation}
\frac{\sqrt{x^{2}+z^{2}}}{a}\approx \frac{v_{D_0}}{a\omega _{E}}=\frac{\left(
c/23\right) ^{2}}{R_{0}\omega _{c_0}}\frac{B_{0}}{E_{a}}\sim \frac{a}{R_{0}}%
\frac{3.5\text{ MeV}}{eE_{a}a}  \label{extalp}
\end{equation}
The previous analysis of Secs.~\ref{Sec:II}, \ref{Sec:III}, \ref{Sec:IV}, and \ref{Sec:V}, based on a comparison with a tokamak
configuration providing the same confinement properties of the thermal
orbits, leads to a value of the voltage drop between the edge and the
center, $E_{a}a$, in between several hundreds and one thousand of kV; with
these values, according to Eq.$\left( \text{\ref{extalp}}\right) $, alpha
particles appear to be confined, $v_{D_0}/\omega _{E}<a$, during their
thermalization process in a WDRT.

\section{Wave-particle interaction}

\label{Sec:appC}

Linear and angular momentum transfers from waves to particles is a central
problem of plasma physics. The transfer process can be collisionless or
collisional and either the orbital or the intrinsic~\cite{Shvets2002,Kostyukov2002}
angular momentum of the wave can be used. Her we restrict the analysis to
collisionless orbital angular momentum transfer which can be reduced to the
analysis of linear momentum transfer along and across the field lines.
Across the field lines, the linear wave momentum generates a charge
separation source of an electric field, this electric field drives a plasma
flow so that the wave-particle momentum balance is satisfied. Consider a
particle with mass $m$ and charge $e$ in an homogeneous static magnetic
field ${\bf B}=B_{0}$ ${\bf e}_{z}$ directed along the $z$ axis of a
Cartesian basis $\left( {\bf e}_{x},{\bf e}_{y},{\bf e}_{z}\right) $. The
particle position is ${\bf r}=x$ ${\bf e}_{x}+y$ ${\bf e}_{y}+z$ ${\bf e}%
_{z} $ and ${\bf v}$ is the velocity. The magnetic field ${\bf B}=B_{0}$ $%
{\bf e}_{z}$ is described by a vector potential ${\bf A}_{0}=B_{0}x$ ${\bf e}%
_{y}$. We define the canonical momentum as ${\bf p}$ $=$ $m{\bf v+}e{\bf A}%
$ $=$ $p_{x}$ ${\bf e}_{x}$ $+ p_{y}$ ${\bf e}_{y}$ $+P$ ${\bf e}_{z}$.
In addition to the static field, the particle interact with an
electromagnetic wave described by the vector potential ${\bf a}=a\sin
(k_{\bot }x+$ $k_{{\bf \Vert }}z-\omega t)$ ${\bf e}_{y}$, so the full
vector potential of the field configuration is, 
\begin{equation}
{\bf A}={\bf A}_{0}+{\bf a}=B_{0}x\text{ }{\bf e}_{y}+a\sin (k_{\bot }x+k_{%
{\bf \Vert }}z-\omega t)\text{ }{\bf e}_{y}\text{.}
\end{equation}
In order to simplify the analysis, we will take $m\ $as the unit of mass, $%
e^{2}B_{0}/m$ as the unit of electric current and $m/eB_{0}$ as the unit of
time. Thus we switch from the IS legal system $\left[ kg,s,A\right] $ to $%
\left[ m,m/eB_{0},e^{2}B_{0}/m\right] $. The Hamiltonian $H$ of the charged
particle, 
\begin{equation}
H=\frac{1}{2}\left( {\bf p-A}_{0}-{\bf a}\right) ^{2}=\frac{1}{2}\left( {\bf %
p-A}_{0}\right) ^{2}-\left( {\bf p-A}_{0}\right) \cdot {\bf a+}\frac{1}{2}%
{\bf a}^{2}\text{,}
\end{equation}
is then restricted to the unperturbed kinetic energy $\left( 
{\bf p-A}_{0}\right) ^{2}/2$ plus the dipolar term $\left( {\bf p-A}%
_{0}\right) \cdot {\bf a}$ and the ponderomotive term ${\bf a}^{2}/2$ is
neglected.

Then, we use a set of canonical variables displaying the guiding center
position $\left( X_{g},Y_{g}\right) $, and perform a canonical transform
from the old momentum $\left( p_{x},p_{y}\right) $and positions $\left(
x,y\right) $ to the new momentum $\left( I,Y_{g}\right) $and positions $%
\left( \theta ,X_{g}\right) $, such that, 
\begin{equation}
{\bf r}=\left(X_{g}+\sqrt{2I}\sin \theta\right) {\bf e}_{x}+\left(Y_{g}+\sqrt{2I}\cos \theta\right) 
{\bf e}_{y}+z{\bf e}_{z}\text{.}
\end{equation}
The substitution of the new canonical variables in the expression of the
momentum, 
\begin{equation}
{\bf p}={\bf A}_{0}+\sqrt{2I}\cos \theta {\bf e}_{x}-\sqrt{2I}\sin \theta 
{\bf e}_{y}+P{\bf e}_{z}\text{,}
\end{equation}
leads to the angle/action perturbed Hamiltonian: 
\begin{equation}
H=I+\frac{P^{2}}{2}-a\sqrt{2I}\sin \theta \sin (k_{\bot }X_{g}+k_{\bot }%
\sqrt{2I}\sin \theta +k_{{\bf \Vert }}z-\omega t)\text{.}
\end{equation}
In order to expand the wave part of this Hamiltonian as a sum of resonant
interactions we use the classical Bessel expansion $\sin \left( u+b\sin
\alpha \right) $ = $\sum_{l}J_{l}\left( b\right) $ $\sin \left( u+l\alpha
\right) $, followed by the use of the relation: $2J_{l}^{\prime }\left(
b\right) $ = $J_{l-1}\left( b\right) $ $-$ $J_{l+1}\left( b\right) $, 
\begin{equation}
H=I+\frac{P^{2}}{2}-a\sqrt{2I}\sum\limits_{l=-\infty }^{l=+\infty
}J_{l}^{\prime }\left( k_{\bot }\sqrt{2I}\right) \cos \left( l\theta
+k_{\bot }X_{g}+k_{{\bf \Vert }}z-\omega t\right) \text{.}
\end{equation}
For a set of initial conditions, the dynamics is dominated by the nearest
resonance $l=N$, so we have Hamilton equations, 
\begin{eqnarray}
\frac{dY_{g}}{dt} &=&k_{\bot }a\sqrt{2I}J_{N}^{\prime }\left( k_{\bot }\sqrt{%
2I}\right) \sin \left( \omega t+N\theta +k_{\bot }X_{g}\right) , \label{Eq:C7}\\
\frac{dI}{dt} &=&Na\sqrt{2I}J_{N}^{\prime }\left( k_{\bot }\sqrt{2I}\right)
\sin \left( \omega t+N\theta +k_{\bot }X_{g}\right) , \label{Eq:C8}\\
\frac{dP}{dt} &=&k_{\Vert }a\sqrt{2I}J_{N}^{\prime }\left( k_{\bot }\sqrt{2I}%
\right) \sin \left( \omega t+N\theta +kX_{g}\right) , \label{Eq:C9}\\
\frac{dH}{dt} &=&\omega a\sqrt{2I}J_{N}^{\prime }\left( k_{\bot }\sqrt{2I}%
\right) \sin \left( \omega t+N\theta +k_{\bot }X_{g}\right) \label{Eq:C10}\text{.}
\end{eqnarray}

These are Eqs. $\left( \text{\ref{hh}}\right) $, $\left( \text{\ref{hh4}}%
\right) $ and $\left( \text{\ref{hh7}}\right) $, used in Sec.\ref{Sec:V}, in
parametric form ($t$ is the parameter) and expressed with the $\left[
m,m/eB_{0},e^{2}B_{0}/m\right] $ system of units ($\sqrt{2I}=v_{c}/\omega
_{c}$, $Y_{g}=\omega _{c}y_{g}/v_{c}...$). The full development of
quasilinear theory~\cite{Stix1992}, where these relations are considered within the
framework of the random phase approximation above a stochasticity threshold,
ultimately leads to the very same conclusion which can be expressed in terms
of diffusion paths in the action space $\left( I,P,Y_{g}\right) $. The
equations of these diffusion paths being given by Eqs. $\left( \text{\ref{hh}%
}\right) $, $\left( \text{\ref{hh4}}\right) $ and $\left( \text{\ref{hh7}}%
\right) $. Equivalent relations have also been derived  \cite{chen88,dodin_08_dp}. In any event, of interest here is that the guiding center displacement $dY_{g}$ represents a flow a space charge
in the perpendicular direction, leading to the radial electric field.
%electric field which induces a plasma flow along the $k_{\bot }$ direction.

%\section{Horizontal polarization, inertial drift and alpha orbits}

\section{Horizontal polarization}

\label{Sec:appD}

The sign of the horizontal magnetic shift $\delta _{B}$, given in %  Eq. $\left( \text{\ref{db}}\right) $, 
Eq.~(\ref{db}),
is the product of the sign of $\omega _{c}$
times the sign of $v_{\Vert }$. 
Thus, for isotropic ion and electron populations, there is no horizontal polarization of the plasma in a tokamak.
The situation is different for the sign of the horizontal electric shift $\delta _{E}$,  Eq.~(\ref{de}), which is given by the
sign of $\omega _{c}$. Let us note the sign of the horizontal polarization.   
For a positively charged torus, where the electric field points outward from the magnetic axis, ions will be shifted horizontally towards the high-field side torus, while electrons will be shifted horizontally towards the low-field side torus.  For a negatively charged torus, the opposite occurs, with the ions  shifted outward in major radius, and the electrons inward.
Equivalently, consider two horizontal cylinders with the same axis and the same radius:
one is filled with hydrogen ions of charge $e$ and density $n$ and the other
with electrons of charge $-e$ and density $n$. 
If the two axes, while remaining parallel and in the same horizontal plane, are slightly separated
by a distance $\delta $, an internal horizontal uniform electric field of
strength $ne\delta /\varepsilon _{0}$ is generated and an external dipolar
field appears.

Thus, even if $\delta _{E}\ll a$, a horizontal separation of charges takes place, leading to a weak horizontal polarization.  
The horizontal polarization leads to a vertical $E \times B$ drift, in which both electrons and ions together drift upward or downward out of the torus.To avoid this electric polarization, the edge space charge, due to the sign dependance of $\delta _{E}$, needs to be short
circuited in a WDRT trap.

However, since the horizontal drift  is weak, the rotational transform that stabilizes the vertical drifts will also stabilize the horizontal drifts.  
To the extent that any residual polarization remains, there exist several possibilities to counteract those polarizations, or possibly even utilize the polarizations for positive purposes.

%\textcolor{red}{  How big is this? If it just leads to vertical drift, then so what if rotational transform is big enough?  How come it doesn't just add to the grad B cross B drift, and then get compensated in the same way? }

One, consider that the upper bound of the thickness of this edge space charge is of the
order the fast ion Larmor radius.  This allows charge to be scraped off the peripheral flux surface.  
In fact, this task can be performed simultaneously with particle and power extraction at the edge. 
Also, if a very small amount of internal, or external, toroidal current is driven to define non degenerate
magnetic surfaces adjusted to iso-potential surfaces, a divertor X point can
be created and the associated SOL thickness can be adjusted to $\delta _{E}$
to solve both polarization and power extraction problems at the same time.

Two, and related to the first possibility, consider that the orbits which give rise to the horizontal field $ne\delta /\varepsilon_{0}$ are edge orbits ($r\sim a$).
As such,  they correspond to the cold part of the discharge and can be intercepted by left and right limiters which can be
connected externally to short circuit this polarization.
 A small amount of edge poloidal magnetic field can also provide a short circuit path along the helicoidal field lines to cancel horizontal polarization.

Three, it is also possible to counteract the polarization with a wave-driven horizontal current. The efficiency of the perpendicular current generation is larger than that of parallel current generation, so that the power cost of
this additional wave driven current will be small.

Thus, horizontal polarization can be cancelled through a stripping of the cold edge orbits with an X point/SOL divertor configuration or with limiters, or short circuited with wave-driven internal horizontal or poloidal currents. 
Note that the external draining of the space charge with limiters or a SOL provides a direct energy recovery process, whereas the internal wave-driven short circuit requires additional power.
On the other hand, the steady-state stripping of the edge cold orbits is in any event mandatory for all toroidal thermonuclear traps, even without horizontal polarization, to remove fusion ash and heat.
Thus some form of stripping of the edge orbits is likely to be the best way to short-circuit the horizontal polarization.

%Wong Effects of Ion-Cyclotron Harmonic Damping on Current Drive in The Lower Hybrid Frequency-Range \cite{wong}
%Parallel rf Force Driven by the Inhomogeneity of Power Absorption in Magnetized Plasma \cite{Gao13}
%F and Rax  Current drive by lower hybrid waves in the presence of energetic alpha particles \cite{fisch92b}

%\bibliography{aipsamp}
%\bibliography{ReferencesRotatingTokamak,more_electric}

%  \bibliography{ReferencesRotatingTokamak}


\end{document}